\documentclass[11pt]{article} 
\usepackage{amsfonts,cite}
\makeatletter \renewcommand{\section}{\@startsection {section}{1}{\z@}
{-3.5ex plus -1ex minus -.2ex}{2.3ex plus .2ex}{\large\bf}}
\renewcommand{\subsection}{\@startsection{subsection}{2}{\z@} {-3.25ex
plus -1ex minus -.2ex}{1.5ex plus .2ex}{\normalsize\bf}}
\makeatletter \@addtoreset{equation}{section} \makeatother

\newcommand{\qh}{\hat{q}} \newcommand{\hh}{\hat{h}}
\newcommand{\hH}{\hat{H}} \newcommand{\hf}{\hat{f}}
\newcommand{\nn}{\nonumber}  \newcommand{\spa}{\,,\qquad}
\newcommand{\gym}{g_{\mathrm{YM}}}

\newcommand{\eqref}[1]{(\ref{#1})} 
\newcommand{\ZZ}{\mathbb{Z}}  
\newcommand{\IR}{\mathbb{R}} \newcommand{\IC}{\mathbb{C}} 
\begin{document}

\renewcommand{\thefootnote}{\fnsymbol{footnote}}
\setcounter{footnote}{1}

\begin{titlepage}

\vspace*{-10ex}

\rightline{\vbox{\footnotesize\hbox{CERN-TH/2001-359}
\hbox{HUTP-01/A075} \hbox{NORDITA-2002/11 HE} \hbox{SPIN-2002/06}
\hbox{ITP-UU-02/07} \hbox{ITFA-2002-05}\vskip.5ex 
\hbox{\tt hep-th/0203064}}}

\vspace{7ex}

\begin{center}

{\LARGE \bf Non-extremal fractional branes\footnote{Work supported in
part by the European Community's Human Potential Programme under
contract  HPRN-CT-2000-00131 Quantum Spacetime.}}

\vspace{8ex}

{\large \bf M. Bertolini${}^a$, T. Harmark${}^{b}$,
N.A. Obers${}^{c,d,e}$  and A. Westerberg${}^{f}$}

\vspace{4ex}

{\it ${}^a$NORDITA, Blegdamsvej 17, DK-2100 Copenhagen \O,
Denmark\\[.75ex] ${}^b$Jefferson Physical Laboratory, Harvard
University, Cambridge, MA 02138, USA \\[.75ex] ${}^c$Spinoza Institute
and Institute for Theoretical Physics,\\[-.3ex] Utrecht University,
Leuvenlaan 4, 3584 CE Utrecht, The Netherlands \\[.75ex]
${}^d$Institute for Theoretical Physics, University of Amsterdam
\\[-.3ex] Valckenierstraat 65, 1018 XE Amsterdam, The Netherlands
\\[.75ex] $^e$The Niels Bohr Institute, Blegdamsvej 17, DK-2100 
Copenhagen \O, Denmark
\\[.75ex] ${}^f$CERN, TH Division, CH-1211 Geneva 23, Switzerland}

\vspace*{3ex}

{\small\tt teobert@nordita.dk, harmark@bose.harvard.edu,\\[-.75ex]
N.Obers@phys.uu.nl, Anders.Westerberg@cern.ch}

\end{center}

\vspace{1ex}

\begin{abstract}
We construct non-extremal fractional D-brane solutions of type-II
string theory at the $\ZZ_2$ orbifold point of K3. These solutions
generalize known extremal fractional-brane solutions and provide
further insights into $\mathcal{N}=2$ supersymmetric gauge theories
and dual descriptions thereof. In particular, we find that for these
solutions the horizon radius cannot exceed the non-extremal
enhan\c{c}on radius. As a consequence, we conclude that a system of
non-extremal fractional branes cannot develop into a black brane. This
conclusion is in agreement with known dual descriptions of the system.
\end{abstract}

\vspace*{3ex}

\end{titlepage}

\renewcommand{\thefootnote}{\arabic{footnote}}
\setcounter{footnote}{0} \setcounter{page}{2} \setcounter{tocdepth}{2}

%%%%%%%%%%%%%%%%%%%%%%%%%%%%%%%%%%%%%%%%%%%%%%%%%%%%%%%%%%%%%%%%%%%%%%%%%%%

\tableofcontents

%%%%%%%%%%%%%%%%%%%%%%%%%%%%%%%%%%%%%%%%%%%%%%%%%%%%%%%%%%%%%%%%%%%%%%%%%%%

%%%%%%%%%%%%%%%%%%%%%%%%%%%%%%%%%%%%%%%%%%%%%%%%%%%%%%%%%%%%%%%%%%%%%%%%%%%
\vspace{1ex}
\section{Introduction}

Fractional D-branes
\cite{Douglas:1997xg,Douglas:1997de,Polchinski:1997ry} have proved an
interesting and rich subject in string theory, generalizing the ordinary
D-branes of type-II string theory. They differ from the latter not only
because they carry fractional charges but also in that their world-volume
gauge theories are in general non-conformal. In particular, fractional
D-branes with eight supercharges can be obtained from type-II string
theory on an orbifold limit of K3, the simplest one being $T^4\!/\ZZ_2$,
or on the orbifold limit of an ALE space, which is $\IC^2\!/\Gamma$ where
$\Gamma$ corresponds to the ADE-classified finite symmetry group of the
ALE space~\cite{Douglas:1996sw,Johnson:1997py}.

In the present paper we consider the type-II string theories on the
$T^4\!/\ZZ_2$ orbifold limit of K3. For these cases, fractional
D-branes have half the charge of the usual regular D-branes. Their
world-volume theories contain only a vector multiplet, while the
regular branes carry in addition a hypermultiplet. The extremal
supergravity solutions for this class of fractional branes were found
in ref.~\cite{Frau:2000gk}. Here we study the non-extremal
generalizations of these solutions.

The three main reasons for studying this problem are as follows. The
first is that it is interesting to determine whether fractional branes
in string theory are dynamical objects in the same sense as ordinary
D-branes. Making the branes non-extremal tests whether it is possible
to consider thermally excited fractional D-branes.

The second reason is related to the fact that the gauge theories
living on the fractional branes that we consider are 3+1-dimensional
pure $\mathcal{N}=2$ super-Yang--Mills (SYM) theory or dimensional
reductions thereof. These are non-conformal theories and it is of
principal interest to extend the Maldacena
conjecture~\cite{Maldacena:1997re,Gubser:1998bc,Witten:1998qj} to
such settings. To make the fractional branes non-extremal would in
principle enable us to obtain a dual description of finite-temperature
pure SYM theories with eight supercharges.

The study of a supergravity/gauge-theory duality for pure SYM with
eight supercharges was initiated in ref.~\cite{Johnson:1999qt} by
considering D-branes wrapping K3, a setup which is T-dual to the one
with fractional D-branes~\cite{Johnson:1999qt,Karch:1998yv,Dasgupta:1999wx}.
The upshot of the analysis in ref.~\cite{Johnson:1999qt} is that the
supergravity solution dual to pure SYM with eight supercharges has a
repulson singularity~\cite{Behrndt:1995tr,Kallosh:1995yz,Cvetic:1995mx}
near the center. However, this repulson can be excised from the solution
by noticing that at a certain distance from the center, the so-called
\emph{enhan\c{c}on radius}, an abelian field in the effective theory
becomes non-abelian, the gauge symmetry being enhanced from U(1) to
SU(2). This means that the low-energy effective theory contains
additional massless fields which have to be taken into account.
Moreover, a probe computation shows that D-brane probes become massless
as one reaches the enhan\c{c}on radius. The interpretation of this
phenomenon is that there is a spherical distribution of D-branes at
the enhan\c{c}on radius with flat space inside the
sphere~\cite{Johnson:1999qt}. Similar results have subsequently also
been obtained for other configurations of wrapped
branes~\cite{Gauntlett:2001ps,Bigazzi:2001aj,DiVecchia:2001uc} as well
as for fractional brane configurations both on non-compact orbifolds like
$\IC^2\!/\Gamma$~\cite{Bertolini:2000dk,Polchinski:2000mx,Grana:2001xn,
Bertolini:2001qa, Billo:2001vg,Bain:2001kw} and on the compact
$T^4\!/\ZZ_2$ orbifold limit of K3~\cite{Frau:2000gk}.

The third main reason for our interest in non-extremal fractional
branes is the interplay between the non-extremality features and the
enhan\c{c}on mechanism; since extremal fractional branes on $T^4\!/\ZZ_2$ 
have an enhan\c{c}on radius~\cite{Frau:2000gk}, it is of interest to see
whether a horizon covering the enhan\c{c}on can develop and thus provide
another kind of excision mechanism for the fractional branes. For systems
of D6-branes wrapped on K3, this question has been addressed in
refs~\cite{Johnson:1999qt,Johnson:2001wm} where it was found that a
horizon can indeed be formed. However, in the supergravity/gauge-theory
duality the horizon radius corresponds to energies that are beyond reach
in the pure SYM theory. This observation has been regarded as evidence
for the conjecture that the dual of pure SYM with eight supercharges is
a non-gravitational theory~\cite{Johnson:1999qt,Wijnholt:2001us}. As we
are going to show, fractional branes provide further corroboration of this
conjecture.

For $\mathcal{N}=1$ SYM in 3+1 dimensions the thermal version of the
Klebanov--Strassler setup~\cite{Klebanov:2000hb} that builds on the
paper~\cite{Klebanov:1996un} has been explored in
refs~\cite{Buchel:2000ch,Buchel:2001gw,Gubser:2001ri}. Here it has
proved very hard to find an exact non-extremal solution. For the
$\mathcal{N}=1^*$ SYM theory progress has been made in
refs~\cite{Freedman:2000xb,Maldacena:2000yy,Gubser:2001eg,Buchel:2001dg},
building on the setup of ref.~\cite{Polchinski:2000uf} describing
zero-temperature $\mathcal{N}=1^*$ SYM. In the latter case, decoupling
the scale of confinement from the string scale in the NS5-brane
world-volume theory has turned out to be problematic, making it
difficult to do computations. However, the problems encountered in
$\mathcal{N}=1$ and  $\mathcal{N}=1^*$ SYM seem rather unrelated to
the problems of finding a non-extremal dual to $\mathcal{N}=2$ SYM.

The summary and organization of the paper are as follows. In
section~\ref{sect:solutions} we present the supergravity solutions
describing non-extremal fractional branes at low energy. More
precisely, we consider a bound state of $M$ fractional D$p$-branes of
type-II string theories on the orbifold $T^4\!/\ZZ_2$. Our solutions
thus generalize the fractional-brane solutions of
ref.~\cite{Frau:2000gk}. Their structure turns out to be somewhat
simpler than that of those found in analogous investigations of the
$\mathcal{N}=1$ theory in
refs~\cite{Buchel:2000ch,Buchel:2001gw,Gubser:2001ri}. In
section~\ref{sect:enhancon} we discuss the physical properties of the
non-extremal solutions, focusing on the interplay between the
non-extremal version of the enhan\c{c}on and the black-hole horizon
$r_0$. In particular, we find that the horizon radius cannot exceed
the enhan\c{c}on radius. As a consequence, we conclude that these
systems of non-extremal fractional branes cannot develop into black
branes. The consequences of the latter conclusion and further
interpretation of the results are our primary concerns in
section~\ref{disc}. Finally, some computational details are given in
an appendix. There we also discuss another solution branch, with
well-defined black-hole thermodynamics but the physical interpretation
of which is presently not clear.

%%%%%%%%%%%%%%%%%%%%%%%%%%%%%%%%%%%%%%%%%%%%%%%%%%%%%%%%%%%%%%%%%%%%%%%%%%%%
%%%%%%%%%%%%%%%%%%%%%%%%%%%%%%%%%%%%%%%%%%%%%%%%%%%%%%%%%%%%%%%%%%%%%%%%%%%%
\vspace{1ex}
\section{The non-extremal solutions}
\label{sect:solutions}

In this section we present the non-extremal low-energy supergravity
solutions for fractional D$p$-branes on K3 in the $T^4\!/\ZZ_2$
orbifold limit. For the type-IIA case, the relevant truncated
six-dimensional supergravity action was obtained in
ref.~\cite{Frau:2000gk} by compactification of type-IIA supergravity
on $T^4\!/\ZZ_2$, and used to construct extremal solutions for
fractional D0- and D2-branes. After deriving the corresponding model
on the type-IIB side, we will consider the non-extremal generalizations
of these solutions for the full range $p=0,1,2,3$.

%%%%%%%%%%%%%%%%%%%%%%%%%%%%%%%%%%%%%%%%%%%%%%%%%%%%%%%%%%%%%%%%%
\subsection{Fractional branes}

Fractional
branes~\cite{Douglas:1997xg,Douglas:1997de,Polchinski:1997ry} are
certain types of BPS D-branes that one encounters when considering
string theory in singular backgrounds. They can be defined in many
different (but equivalent) ways. Probably, the most intuitive way to
understand their properties is through their description as
higher-dimensional D-branes wrapped on a vanishing cycle of the
singular manifold~\cite{Diaconescu:1998br,Billo:2000yb}. This
geometric picture makes manifest the characteristic feature of
fractional branes as being stuck at the singularity while free to move
in the flat transverse directions only. This viewpoint also shows why
they lack the world-volume degrees of freedom associated to
fluctuations in the orbifold directions which correspond, in general,
to hypermultiplet excitations. Moreover, as will become clear below,
this also automatically accounts for the coupling of fractional branes
with the twisted sector of string theory on the orbifold. We refer to
ref.~\cite{Bertolini:2001gq} for a recent review on the properties of
fractional branes on orbifolds for theories with eight supercharges.

Let us now focus on our main case of interest. The K3 manifold has
${\rm dim}(H_2(\mathrm{K3}))\allowbreak=22$ two-cycles, on a space
of signature (19,3). At the $\ZZ_2$ orbifold point, there are three
self-dual and three anti-selfdual cycles from the six two-cycles in
$H_2(T^4)$, which are invariant under the $\ZZ_2$ involution. In
addition, there are 16 anti-selfdual cycles that come from the
collapsed spheres at the 16 orbifold singularities. A fractional
D$p$-brane is then a D($p$+2)-brane wrapped on one of the these
cycles, $\mathcal{C}$, in the $\ZZ_2$ orbifold limit of K3. The
presence of a background NS-NS two-form flux through the shrinking
cycle makes the D-brane tension non-vanishing. The background value
of this flux is dictated to be
\begin{equation}
\label{bgPhi}
b_0 = \int_{\mathcal{C}} B_{(2)} = \frac{1}{2}(2\pi\sqrt{\alpha'})^2
\end{equation}
by the requirement of conformal invariance of the string world-sheet
in the orbifold background~\cite{Aspinwall:1996mn}. The effective
theory describing the dynamics of such a brane at low energy is a
($p$+1)-dimensional pure SYM theory with eight supercharges. From the
closed string theory point of view, this translates into the fact that
fractional branes couple to both the untwisted and the twisted sector
of string theory on the orbifold.

At low energy, in the infinite-volume limit of $T^4\!/\ZZ_2$, a given
fractional D$p$-brane simply couples to the metric, the dilaton and
the RR ($p$+1)-form potential in the untwisted sector, and to a scalar
field and a ($p$+1)-form potential in the twisted sector. The two latter
correspond, respectively, to the zero mode of the NS-NS $B_{(2)}$ field
and of the RR ($p$+3)-form potential when ``dimensionally reduced'' on
the shrinking two-cycle, and belong to a non-gravitational multiplet of
the effective supergravity theory~\cite{Diaconescu:1998br,Billo:2000yb}.
Denoting by $\omega_{(2)}$ the closed differential two-form Poincar\'e
dual to the vanishing two-cycle $\mathcal{C}$ on which the branes are
wrapped, the relations between higher-dimensional forms and twisted
fields read
\begin{equation}
\label{twistedfields}
B_{(2)} = b \, \omega_{(2)} \spa C_{(p+3)}= \sqrt{2V} A_{(p+1)} \wedge
\omega_{(2)} \,.
\end{equation}
In the compact case (anticipated here by the introduction of the
dimensionful constant $V$ to be defined shortly) the effective
six-dimensional theory is augmented by the zero modes of massless
fields on the internal manifold. However, these come only from the
untwisted sector since, by construction, the twisted fields have no
dynamics on the internal space. For $\omega_{(2)}$ we adopt the
following conventions:
\begin{equation}
\label{omegaconv}
\int_{\mathcal{C}} \omega_{(2)} = 1 \spa \int_{\IC^2\!/\ZZ_2}
{*}\omega_{(2)}\wedge\omega_{(2)} = \frac{1}{2} \spa
\omega_{(2)}+{*}\omega_{(2)} = 0 \,.
\end{equation}

%%%%%%%%%%%%%%%%%%%%%%%%%%%%%%%%%%%%%%%%%%%%%%%%%%%%%%%%%%%%%%%%%%%%%%
\subsection{The actions}
\label{subsect:lagrangian}

Let us start by fixing our conventions and presenting the consistently
truncated six-dimensional actions describing the dynamics of the
supergravity fields to which the fractional D$p$-branes couple. As
just discussed, the gauge fields that enter the solution for a given
$p$ are the two ($p$+1)-form potentials $C_{(p+1)}$ and $A_{(p+1)}$.
{}From the twisted scalar field arising from the NS-NS two-form
potential we separate out a fluctuating part $\tilde b$ according to
\begin{equation}
\label{Vsdef}
b = \frac{1}{2}\sqrt{V_*} + \sqrt{2V}\,\tilde b \spa
V_*=(2\pi\sqrt{\alpha'})^4 \,,
\end{equation}
where $V$ is the volume of the compact space $T^4\!/\ZZ_2$ and $V_*$
is introduced as a shorthand notation. Furthermore, the Kaluza--Klein
reduction of the metric gives rise to four scalar fields,
$e^{2\eta_a},\,a=6,\ldots,9$. Assuming homogeneity in the compactified
directions these are equal and can be replaced by a single scalar field
defined as $\eta = \sum_{a=6}^9 \eta_a$. The dimensionally reduced
dilaton $\phi=\phi|_{d=10}-\frac{1}{2}\eta$ completes the scalar field
content.

For the case $p=0$ the truncated bosonic $d=6$ action governing the
dynamics of these fields was derived in ref.~\cite{Frau:2000gk},
together with the electro-magnetically dual two-brane case. From these
results the corresponding action for the fractional D3-brane on the
type-IIB side can readily be inferred. These three actions take the
form\footnote{Our conventions are ${*}\,\xi_{(6-n)}=\frac{1}{n!}
dx^{\mu_1}\wedge\ldots\wedge dx^{\mu_n}\,\epsilon_{\mu_1\ldots\mu_n}
{}^{\mu_{n+1}\ldots\mu_{6}}\,\xi_{\mu_{n+1}\ldots\mu_6}$ and
$\epsilon^{012345}=+1$. The RR field strengths in ten dimensions were
defined as
$\hat{G}_{(n+1)}=d\hat{C}_{(n)}+\hat{H}_{(3)}\wedge\hat{C}_{(n-3)}$.}
\begin{eqnarray}
\label{thelagrangian}
S_{\mathrm{bulk}}^{(p=0,2,3)} &=& \frac{V}{2\kappa^2}\int \Big\{
 d^6x\sqrt{-g}R + \,{*}d\phi \wedge d\phi +
 \frac{1}{4}\,{*}d\eta\wedge d\eta
 +\frac{1}{2}e^{-\eta}\,{*}d\tilde{b}\wedge d\tilde{b} \nonumber \\ &&
 \qquad\quad
 +\,e^{(1-p)\phi}\Big[\frac{1}{2}e^\eta\,{*}G_{(p+2)}\wedge G_{(p+2)}
 + \frac{1}{2}\,{*}\tilde{F}_{(p+2)}\wedge\tilde{F}_{(p+2)} \Big]
 \Big\} \,,
\end{eqnarray}
where $\kappa = 8 \pi^{7/2} g_s \alpha'^{2}$ is the ten-dimensional
gravitational coupling and
\begin{equation}
\label{GIFstrengths}
G_{(p+2)} = d C_{(p+1)} \spa \tilde{F}_{(p+2)} = d A_{(p+1)} + d
\tilde{b} \wedge C_{(p+1)}
\end{equation}
are the gauge-invariant field strengths of the untwisted- and
twisted-sector  potentials $C_{(p+1)}$ and $A_{(p+1)}$, respectively.

For the remaining case, $p=1$, some subtleties arise because the field
strength $\tilde{F}_{(3)}$ is self-dual, a property inherited from its
ten-dimensional type-IIB five-form parent. As usual, the self-duality
condition has to be imposed on shell. With this proviso in mind we can
nevertheless write a truncated $d=6$ gravity action also for the
fractional D-string:%
\footnote{This action results from taking a type-IIB action in ten
dimensions compatible with an \emph{anti}-selfdual five-form field
strength $\hat{G}_{(5)}$, and using the conventions~\eqref{omegaconv}
in the Kaluza--Klein reduction on $T^4\!/\ZZ_2$.}
\begin{eqnarray}
\label{action1}
S_{\mathrm{bulk}}^{(p=1)} &=& \frac{V}{2\kappa^2} \int \Big\{
 d^6x\sqrt{-g}R + \,{*}d\phi \wedge d\phi +
 \frac{1}{4}\,{*}d\eta\wedge d\eta
 +\frac{1}{2}e^{-\eta}\,{*}d\tilde{b}\wedge d\tilde{b} \nonumber \\ &&
 \quad\quad +\, \frac{1}{2}e^\eta\,{*}G_{(3)}\wedge G_{(3)} +
 \frac{1}{4}\,{*}\tilde{F}_{(3)}\wedge\tilde{F}_{(3)} - \frac{1}{2}
 A_{(2)} \wedge G_{(3)} \wedge d\tilde{b} \Big\} \,.
\end{eqnarray}
The $C_{(2)}$ equation of motion is compatible with the self-duality
constraint ${*}\tilde{F}_{(3)}=\tilde{F}_{(3)}$, a fact which relies
crucially on the presence of the Chern--Simons term.

In solving the equations of motion obtained from the
actions~\eqref{thelagrangian} and~\eqref{action1} one should also
specify the boundary conditions at infinity that the various fields
should satisfy (i.e.\ the mass and the charges of the soliton). We are
interested in describing a general bound state of non-extremal
fractional D$p$-branes whose corresponding supergravity solution
should match the extremal one at infinity. The boundary conditions at
infinity for the latter are encoded in the bosonic world-volume action
describing the low-energy dynamics of an extremal fractional
brane. For the case at hand this action has been derived in
ref.~\cite{Frau:2000gk} and reads
\begin{eqnarray}
\label{bac1}
S_{\mathrm{wv}} &=& - \frac{T_p}{2\kappa} \int d^{p+1}\xi \,
\sqrt{-g}\, e^{-\eta/2} e^{-(1-p)\phi/2} \left( 1 + 2 \sqrt{\frac{2
V}{V_*}} \tilde b \right) \nonumber \\[1ex] && \qquad +
\,\frac{T_p}{2\kappa} \int \left[ C_{(p+1)} \left( 1 + 2 \sqrt{\frac{2
V}{V_*}} \tilde b \right) + 2 \sqrt{\frac{2 V}{V_*}} A_{(p+1)} \right]
\,,
\end{eqnarray}
where $T_p=\sqrt{\pi} (2 \pi \sqrt{\alpha'})^{3-p}$. The world-volume
action for a stack of $M$ coincident fractional D$p$-branes is
obtained simply by multiplying the above action by $M$ (this will be
implicitly assumed in what follows).

%%%%%%%%%%%%%%%%%%%%%%%%%%%%%%%%%%%%%%%%%%%%%%%%%%%%%%%%%%%%%%%%%%%%%%%%
\subsection{Ans\"atze and solutions}
\label{subsect:ansaetze}

Below we present the solutions of the equations of motion, given in
appendices~\ref{subsect:EOMs} and~\ref{subsect:spherical}, referring
to appendix~\ref{subsect:solving} for an outline of their derivation.
We will wherever possible treat the cases $p=0,1,2,3$ in parallel. As
far as the starting-point---i.e.\ the ansatz for the metric---is
concerned, the three-brane (being of codimension two in six dimensions)
however needs some special consideration. The solutions, nevertheless,
share a common structure for all four cases as will become clear below.

Let us first discuss the lower-dimensional cases, $p=0,1,2$, for which
the standard non-extremal $p$-brane ansatz applies:
\begin{equation}
\label{h_ansatz}
ds^2 = H^{\frac{p-3}{4}} \Big( - f dt^2 + \sum_{i=1}^p (dx^i)^2 \Big)
 + H^{\frac{p+1}{4}} \left( f^{-1}dr^2 + r^2 d\Omega_{4-p}^2 \right)\,.
\end{equation}
The non-extremality is introduced by the function $f$ which, like $H$,
depends on the transverse radial coordinate only. It is constrained by
the equations of motion to satisfy the harmonic equation
\begin{equation}
\label{eq:f_eq}
f'' + \frac{4-p}{r}f' = 0 \,.
\end{equation}
Requiring the non-extremal solution to approach the extremal one at
infinity, the solution to eq.~\eqref{eq:f_eq} takes the form
\begin{equation}
\label{f_ansatz}
f = 1 - \left(\frac{r_0}{r}\right)^{3-p} \,,
\end{equation}
with the horizon radius $r_0$ governing the degree of
non-extremality.\footnote{Although the metric~\eqref{h_ansatz}
develops a horizon at $r_0$ we shall see in the next section that the
supergravity approximation ceases to be valid at a radius $r_e$
strictly larger than $r_0$. Keeping this in mind, we shall
nevertheless refer to $r_0$ as the horizon radius.}

The upshot of the analysis of appendix~\ref{subsect:solving} is that
the non-extremal solutions can be expressed entirely in terms of $f$
and two additional harmonic functions, $h_1$ and $h_2$, which in
turn depend on three parameters: the horizon radius $r_0$ and two
charges, $q_1$ and $q_2$, depending linearly on the number of branes
which act as sources for the  solution. More specifically, the scalar
fields are given by
\begin{equation}
\label{scal1}
e^\phi = H^{\frac{1-p}{4}} \spa e^\eta = \frac{H}{h_1^2} \spa \tilde b
= \frac{q_2}{q_1} \left(\frac{h_2}{h_1}-1\right) \,,
\end{equation}
while the gauge potentials take the form
\begin{equation}
\label{potsols}
C_{0\ldots p} = - \frac{q_2}{r^{3-p}} \frac{h_3}{H} \spa A_{0\ldots p}
= - \frac{q_1}{r^{3-p}} \frac{1} {h_1} \,,
\end{equation}
the associated field strengths (defined in eq.~\eqref{GIFstrengths} above) 
being
\begin{equation}
\label{fstrengthsols}
G_{r0\ldots p} =  \frac{(3-p) q_2}{r^{4-p}} \frac{h_1 h_2}{H^2} \spa
\tilde{F}_{r0\ldots p} =  \frac{(3-p) q_1}{r^{4-p}} \frac{1}{H} \,.
\end{equation}
The functions $H$ and $h_3$ entering the solution read
\begin{equation}
\label{Hh3defs}
H = \left( 1 + \frac{1}{2} \frac{q_2^2}{q_1^2} \right) h_1^2 -
\frac{1}{2} \frac{q_2^2}{q_1^2}\, h_2^2 \spa h_3 = \frac{1}{2} (h_1 +
h_2) \,,
\end{equation}
where the two basic harmonic functions are
\begin{equation}
\label{accai}
h_i = 1 -\Big(\frac{r_i}{r}\Big)^{3-p} \spa i =1,2 \,.
\end{equation}
The radial parameters of these functions are given by
\begin{eqnarray}
\label{r1}
r_1^{3-p} &=& \frac{1}{2} r_0^{3-p} + \frac{\sqrt{2q_1^4 + (q_1^2 +
 q_2^2)r_0^{2(3-p)} - 2 q_1^2 \Lambda}}{2\sqrt{2q_1^2+q_2^2}}\,,
 \\[2ex]
\label{r2}
r_2^{3-p} &=& \frac{1}{2} r_0^{3-p} + \frac{\sqrt{2q_1^4 + (q_1^2 +
 q_2^2)r_0^{2(3-p)} + 2 q_1^2 \Lambda}}{2 \sqrt{q_2^2}}\,,
\end{eqnarray}
where
\begin{equation}
\label{lambda}
\Lambda \equiv \sqrt{q_1^4 + (q_1^2 +q_2^2)r_0^{2(3-p)} +
 \frac{1}{4}r_0^{4(3-p)}} \,.
\end{equation}

As already noticed, the solution we have found is completely fixed by
extracting the relation between the free-supergravity values of the
charges $q_1$ and $q_2$ and those dictated by the world-volume
action~\eqref{bac1} for the $M$ fractional branes. Using eqs~\eqref{Q1}
and~\eqref{Q2} and equating the corresponding charges with the coupling
to the twisted and untwisted potentials $A_{0\ldots p}$ and
$C_{0\ldots p}$  in the WZ action one gets
\begin{equation}
\label{q1q2}
q_1 \,=\, \sqrt{\frac{2 V}{V_*}} Q_p\,M \spa q_2 \,=\, Q_p\,
\frac{M}{2} \,,
\end{equation}
where
\begin{equation}
\label{def1}
Q_p = \frac{2}{k_p}\,\frac{T_p \,\kappa}{\Omega_{4-p}\,V} = 
  \frac{1}{k_p\,\Omega_{4-p}}\,\frac{V_*}{V}\,(2\pi\sqrt{\alpha'})^{3-p}\,g_s
  \,,
\end{equation}
with $\Omega_{4-p}$ denoting the volume of the unit $(4{-}p)$-sphere
surrounding the $p$-branes and $k_p=3-p$ for $p<3$ while $k_3=1$.%
\footnote{The solution describing a composite bound state of $M$
fractional and $N$ regular D$p$-branes is identical to the one we have
discussed so far, the only difference being that the untwisted charge
$q_2$ will now be $q_2 = Q_p (\frac{M}{2} + N)$. By taking $M=0$ one
gets $q_1=0$ and $q_2 = Q_p N$, giving back the usual regular brane
solution, with no coupling to twisted fields.}

As a consistency check let us also take the extremal limit, $r_0=0$.
Assuming the charges to be positive, as we will from now on, we obtain
\begin{equation}
\label{exth1h2}
h_1^{\mathrm{extr}} = 1 \spa h_2^{\mathrm{extr}} = 1 -
\frac{q_1^2}{q_2} \frac{1}{r^{3-p}} \spa H_{\mathrm{extr}} = 1 +
\frac{q_2}{r^{3-p}} - \frac{q_1^2}{2r^{2(3-p)}} \,,
\end{equation}
so that the solution of ref.~\cite{Frau:2000gk} is recovered:
\begin{equation}
e^{\eta_{\mathrm{extr}}} = H_{\mathrm{extr}} \spa
\tilde{b}_{\mathrm{extr}} = - \frac{q_1}{r^{3-p}}  = A_{0\ldots
p}^{\mathrm{extr}} \spa C_{0\ldots p}^{\mathrm{extr}} =
H^{-1}_{\mathrm{extr}} - 1 \,.
\end{equation}
Note that the structure of the function $H$ that determines the non-extremal 
metric (together with $f$) and the dilaton is the same as for the extremal
case; while the coefficients of course differ, introducing a
non-extremality parameter gives no terms beyond the $r^{-2(3-p)}$
correction, the latter being the usual fractional brane modification
to the harmonic function governing the regular-brane solution. For the
other fields the non-extremal modifications are somewhat more
intricate. Nevertheless, for our class of non-extremal fractional
branes, these modifications are entirely due to the non-triviality of
the harmonic function $h_1$ for $r_0>0$. In particular, we note that,
contrary to the extremal case, the non-extremal twisted fields
\emph{do} get corrections with respect to their harmonic asymptotic
behavior. The absence of such corrections was taken as input for the
ansatz relevant to extremal fractional branes~\cite{Frau:2000gk}, and
was argued (for the NS-NS twisted scalar $\tilde{b}$) to be a
manifestation of the fact that $\mathcal{N}=2$ SUSY only allows for
one-loop perturbative corrections. The fact that this property ceases
to hold for our non-extremal generalization suggests that the
non-extremality parameter $r_0$ indeed does switch on a temperature in
the system.

Turning to the case $p=3$, the appropriate non-extremal ansatz for the
metric reads~\cite{Lu:1997kg}
\begin{equation}
\label{h_ansatz3}
ds^2 = - f dt^2 + \sum_{i=1}^3 (dx^i)^2 +
 \Big(\frac{\tilde{r}}{r}\Big)^2 H \left( f^{-1}dr^2 + r^2 d\theta^2
 \right) \,,
\end{equation}
where $\tilde{r}$ is an (as yet) undetermined radial parameter. Using
this ansatz, the analysis in appendix~\ref{subsect:solving} can be
done in parallel with the lower-dimensional cases. As a consequence,
the expressions listed above for the metric, the scalars and the gauge
field strengths (upon substituting $(3-p)\mapsto1$) in terms of the
functions $h_1$ and $h_2$ are valid also for $p=3$. Although the
reason is slightly less obvious, by using the mapping
$r^{-(3-p)}\mapsto \log(r_\Lambda/r)$ the same turns out to be the
case for the gauge potentials in~\eqref{potsols}, giving
\begin{equation}
\label{potentials3}
C_{0123} = - q_2\,\frac{h_3}{H}\,\log\frac{r_\Lambda}{r} \spa
A_{0123} = - \frac{q_1}{h_1}\,\log\frac{r_\Lambda}{r} \,.
\end{equation}
Here $r_\Lambda$ can be interpreted as a long-distance cut-off in the
transverse radial direction of the three-brane world volume,
corresponding to a UV cut-off on the dual gauge-theory side. This
mapping originates from the harmonic functions which in two dimensions
involve a logarithm. The function $f$, for instance, again satisfies
the harmonic equation~\eqref{eq:f_eq} but the solution now reads
\begin{equation}
\label{fresult3}
f= 1 - a_0 \log\frac{r_\Lambda}{r} \,.
\end{equation}
The dimensionless non-negative parameter $a_0$ governs the degree of
non-extremality. As will become evident below, $a_0$ can be viewed as
the direct formal analogue of the parameter $r_0$ in the solutions for
the lower-dimensional branes. The ``horizon radius''\footnote{We write
the term within quotes since the transverse space is two-dimensional
and black holes therefore can never develop for $p=3$.} for the
three-brane is $r_0=r_\Lambda e^{-1/a_0}$ and the extremal limit is
obtained by letting $a_0\rightarrow0$ with $r_\Lambda$
fixed. Moreover, it is convenient to identify the parameter
$\tilde{r}$ in~\eqref{h_ansatz3} with $r_\Lambda$ since the latter
sets the length scale of the transverse geometry.

Hence, the conditions that $h_1$ and $h_2$ be harmonic now imply
\begin{equation}
\label{h1h2p3}
h_1 = 1 - a_1 \log \frac{r_\Lambda}{r}\,, \qquad h_2 = 1 - a_2 \log
\frac{r_\Lambda}{r}\,.
\end{equation}
In an exact analogy with the results~\eqref{r1} and~\eqref{r2} for
$p<3$ the parameters $a_1$ and $a_2$ are given by
\begin{eqnarray}
\label{a1} a_1 &=& \frac{1}{2} a_0 + \frac{\sqrt{2q_1^4 + (q_1^2 + q_2^2)a_0^2 -
 2 q_1^2 \Lambda}}{2\sqrt{2q_1^2+q_2^2}} \,,\\[1.5ex] \label{a2} a_2
 &=& \frac{1}{2} a_0 + \frac{\sqrt{2q_1^4 + (q_1^2 + q_2^2)a_0^2 + 2
 q_1^2 \Lambda}}{2\sqrt{q_2^2}} \,,
\end{eqnarray}
with
\begin{equation}
\Lambda = \sqrt{q_1^4 + (q_1^2 +q_2^2)a_0^2 + \frac{1}{4}a_0^4} \,.
\end{equation}
The reason for these close formal similarities between $p=3$ and the
lower-dimensional cases is explained in section~\ref{subsect:solving}.

Taking the extremal limit, the gauge potentials~\eqref{potentials3}
simplify to
\begin{equation}
C_{0123}^{\mathrm{extr}} = H^{-1}_{\mathrm{extr}} - 1 \, \spa
A_{0123}^{\mathrm{extr}} = - q_1\,\log\frac{r_\Lambda}{r} \,,
\end{equation}
where $H_{\mathrm{extr}}=1+q_2\log\frac{r_\Lambda}{r}-\frac{1}{2}q_1^2
\left(\log\frac{r_\Lambda}{r}\right)^2$. In addition we have $\tilde
b_{\mathrm{extr}} = -q_1\log\frac{r_\Lambda}{r}$, with the remaining
fields formally identical to their lower-dimensional counterparts.
Note that in the extremal limit of the three-brane solution we have
$\phi=-\frac{1}{2}\eta$. From the definition of $\phi$ this relation
immediately implies that the ten-dimensional dilaton is constant, in
agreement with the fractional D3-brane solution of
refs~\cite{Bertolini:2000dk,Polchinski:2000mx} for the non-compact
orbifold spacetime $\IR^{1,5}\times\IC^2\!/\ZZ_2$.

%%%%%%%%%%%%%%%%%%%%%%%%%%%%%%%%%%%%%%%%%%%%%%%%%%%%%%%%%%%%%%%%%%%%%%%%%%%%%%%
%%%%%%%%%%%%%%%%%%%%%%%%%%%%%%%%%%%%%%%%%%%%%%%%%%%%%%%%%%%%%%%%%%%%%%%%%%%%%%%
\vspace{1ex}
\section{Enhan\c{c}on versus horizon}
\label{sect:enhancon}

In this section we first review the enhan\c{c}on mechanism and
describe its manifestation in the case at hand. Then we examine the
interplay between the event horizon and the enhan\c{c}on shell.

\subsection{Review of the enhan\c{c}on}

Supergravity solutions of brane configurations which have pure SYM
with eight supercharges as their low-energy world-volume theories are
in general plagued by naked singularities. This is in particular true
for fractional branes and, more generally, for D-branes wrapped on
topologically non-trivial cycles. The naked singularities one
encounters are of repulson
type~\cite{Behrndt:1995tr,Cvetic:1995mx,Kallosh:1995yz} and can be
excised by the so-called enhan\c{c}on mechanism~\cite{Johnson:1999qt}.

The logic of this mechanism is as follows. Far away from the source we 
have a perfectly valid supergravity solution. However, when approaching 
the source there is a certain radius $r_e$, dubbed the enhan\c{c}on radius, 
at which the effective supergravity description needs to be augmented by 
additional massless degrees of freedom, the appearance of which leads to 
an enhancement of the gauge symmetry from U(1) to SU(2).\footnote{While 
on the type-IIA side the symmetry enhancement occurs for an ordinary 
vector field, the enhanced $A_1$ gauge symmetry in the type-IIB case 
is carried by a self-dual two-form potential~\cite{Witten:1995zh}.}
A brane-probe calculation shows that the tension of the probe vanishes 
precisely at the enhan\c{c}on radius. The interpretation of this phenomenon 
is that the branes are not located at the origin but instead smeared over
the spherical shell $r=r_e$. The geometry for $r<r_e$ is consequently 
completely different, and, in particular, the singularity is
excised~\cite{Johnson:1999qt,Johnson:2001wm}.

For fractional branes, the symmetry enhancement occurs when the NS-NS 
two-form flux through the vanishing cycle on which the brane is wrapped
flows from the value $1/2$ at infinity to $0$ at the enhan\c{c}on radius; 
when the flux vanishes all parameters of the cycle are zero, which 
corresponds to a point of enhanced symmetry in the moduli space. 
At this point, the fractional branes become tensionless. In particular,
the massless degrees of freedom responsible for the enhanced gauge
symmetry correspond to tensionless fractional D-particles for type IIA
and to tensionless fractional D-strings for type IIB~\cite{Johnson:2000ch}.

For the extremal fractional D-branes on $T^4\!/\ZZ_2$ the enhan\c{c}on
mechanism has been examined in ref.~\cite{Frau:2000gk}. As we are
going to discuss in the next section, the presence of the enhan\c{c}on
has crucial consequences for the decoupling limit and for the nature
of the gauge-theory/gravity duality for this kind of systems. It is
therefore important to investigate if and how the enhan\c{c}on is
modified in the non-extremal case we are discussing, and more
specifically what the relation between the event horizon and the
enhan\c{c}on is.

As discussed in the previous section, fractional D$p$-branes are a
particular kind of wrapped D($p$+2)-branes arising in orbifold
compactifications of string theory. While the geometric volume of the
compact two-cycle characterizing the orbifold is identically zero,
there is a non-trivial NS-NS two-form background flux on it, displayed
in~\eqref{bgPhi}, which makes the effective stringy volume
asymptotically non-vanishing. In fact, as already explained in the
previous section, this asymptotic value is modified by the presence of
the fractional branes (either extremal or non-extremal). Let us recall
the relation between the fields entering the solution discussed in
section~\ref{sect:solutions} and the NS-NS two-form flux:
\begin{equation}
b = \int_{\mathcal{C}} B_{(2)} = \frac{1}{2} \,\sqrt{V_*} +
\sqrt{2V}\, \tilde b\,(r) \,.
\end{equation}
Here we recall that $V$ is the volume of the compact orbifold and
$V^*=(2\pi\sqrt{\alpha'})^4$ as defined in \eqref{Vsdef}. {}From the
above considerations it follows that the enhan\c{c}on is located at
the radius $r_e$ determined by
\begin{equation}
\label{eneq1}
1 + 2 \sqrt{\frac{2 V}{V_*}} \tilde b\,(r_e) = 0 \,.
\end{equation}
This equation gives the position of the enhan\c{c}on in the general
case. In the extremal limit it reduces to the result found in
ref.~\cite{Frau:2000gk}.

Keeping in mind that it is eq.~\eqref{eneq1} that \emph{defines} the
enhan\c{c}on shell, we can also check that fractional D$p$-brane probes
indeed become tensionless there. One simply takes the DBI action for the
probe evaluated in the background generated by the source branes, chooses
static gauge with the transverse coordinates depending on time only, and
finally expands the action up to terms quadratic in the velocities. The
appropriate DBI action is
\begin{equation}
\label{dbi1}
S_{\mathrm{DBI}} = - \frac{T_p}{2\,\kappa} \int d^{p+1} \xi \;
\sqrt{-g}\, e^{-\eta/2} e^{- (1-p) \phi/2} \bigg( 1 + 2 \sqrt{\frac{2
V}{V_*}} \tilde b \bigg) \,.
\end{equation}
By proceeding as just outlined one easily sees that the probe brane becomes
tensionless when eq.~\eqref{eneq1} is satisfied. More precisely, for slow
radial motion ($\dot{r}^2\ll f^2/H$) the kinetic part of its effective
lagrangian is
\begin{eqnarray}
T(r,\dot r) &=& \frac{T_p V_p}{4\kappa}\,\frac{h_1(r)}{f^{3/2}(r)}
  \bigg(1 + 2\sqrt{\frac{2 V}{V_*}} \tilde{b}(r)\bigg) \dot{r}^2 \,,
\end{eqnarray}
showing that the effective tension of the probe is $r$-dependent and that,
as promised, the brane becomes tensionless at the distance $r_e$ given by
eq.~\eqref{eneq1}.%
\footnote{This argument rests on $r_e$ being larger than both
$r_0$ and $r_1$, which, as we shall see, is always the case.}

\subsection{Application to non-extremal fractional branes}

For our class of non-extremal fractional branes, we can obtain a very
simple expression for the enhan\c{c}on. By substituting the solution
discussed in the previous section, in particular eqs~\eqref{scal1} and
\eqref{accai}, in eq.~\eqref{eneq1} we find
\begin{equation}
\label{enhan1}
r_e^{3-p}\,=\, 2 \sqrt{\frac{2 V}{V_*}} \frac{q_2}{q_1}
\left(r_2^{3-p}- r_1^{3-p}\right)+ r_1^{3-p} \,= \,r_2^{3-p} \,,
\end{equation}
where in the last step we have used the relations~\eqref{q1q2}. A
completely analogous expression holds for $a_e$, and thus for $r_e$,
in the three-brane case (recall that $r_e=r_\Lambda e^{-1/a_e}$).%
\footnote{As already mentioned, black holes cannot appear in 2+1
dimensions  so we exclude the three-brane case from the following
discussion.}

We now turn to examining the position of the enhan\c{c}on shell
relative to that of the event horizon. Using eqs~\eqref{r2}
and~\eqref{lambda}, it is easy to see from eq.~\eqref{enhan1} that
the enhan\c{c}on always lies \emph{outside} the horizon, no matter
the value of $r_0$. This means that in the region of validity of the
supergravity approximation, the bound state never develops into a
black brane. This might seem puzzling, since one would think that for
large enough mass the system would indeed develop into a black brane,
while the above equations show that the enhan\c{c}on increases with
$r_0$ faster than $r_0$ itself. However, one has to remember that the
energy density of this configuration is not concentrated in the center
of the solution, but rather spread out on the enhan\c{c}on
shell. Indeed, the fact that we cannot arrange for the horizon to lie
outside the enhan\c{c}on shell is nicely consistent with the fact
that, while the mass is not bounded, the density of mass is. To see
this, note first that the density of mass is the total mass $M_p$
divided by the volume $V_{\rm tot}$ that we can fit the system
into. The mass $M_p$ goes like $r_0^{3-p}$ while the volume $V_{\rm
tot}$ goes like $r_e^{5-p}$, and hence $M_p/V_{\rm tot}$ actually
decreases as we increase $r_0$, since $r_0$ is smaller than $r_e$.

One can also try to extend the solution to the interior of the
enhan\c{c}on shell. Fractional branes cannot get inside the
enhan\c{c}on since their tension vanishes there and their energy
(and charge) is believed to be distributed on the enhan\c{c}on shell.
Therefore, the extremal solution is flat in the interior and the energy
density vanishes
there~\cite{Johnson:1999qt,Merlatti:2001gd}\footnote{By studying the
Seiberg--Witten curve it has been explicitly shown that the
supergravity fields should be constant in the
interior~\cite{Petrini:2001fk}. On the other hand, to discuss the
system at the enhan\c{c}on scale one should include the extra massless
fields into the analysis.}. However, by making the solution
non-extremal one could imagine creating a neutral black hole on the
inside, characterized by an internal horizon radius $r_0'$. One could
then try to increase $r_0'$ enough to make it larger than the
enhan\c{c}on, thus allowing the system to develop into a black brane.%
\footnote{Obviously, we would then have to jump to another branch of
the solution. In terms of the four branches of the solution summarized
in table~\ref{tabbranches} in section~\ref{subsect:iquattrorami} we
should jump from branch I to branch IV.}

If we demand that such an interior black hole be in equilibrium with
the branes at the enhan\c{c}on shell, we can in principle find $r_0'$
in terms of $r_0$ and $M$. This raises the interesting question
whether the interior horizon can reach the enhan\c{c}on
radius. Unfortunately, we cannot answer this question in a precise way
since we do not understand non-extremal fractional branes sufficiently
well to determine when we have thermodynamical equilibrium. However,
the fact that the density $M/V_{\rm tot}$, as defined above, decreases
for increasing $r_0$ makes it seem unlikely that by increasing $r_0$
and thereby decreasing the density of mass $M/V_{\rm tot}$, one could
make the system collapse into a black hole. In the next section we
interpret the fact that the non-extremal fractional-brane system on
the orbifold under study never collapses into a black hole from the
perspective of the SYM theory living on the brane and its supergravity
dual.

At this point we would like to alert the reader to an apparent puzzle.%
\footnote{For simplicity of notation we give expressions appropriate
for the two-brane case although the discussion applies generally.}
The discussion so far can readily be extended to the more general
bound states where $N$ regular branes are also present. As noticed in
the previous section, the only modification of the solution is that we
now have $q_2=Q_2 (M/2 + N)$. Of course, probing the geometry with
regular branes does not give any information on the enhan\c{c}on locus
since regular branes do not couple to the $B_{(2)}$-flux and are
insensitive to the enhan\c{c}on. On the other hand, revisiting the
fractional-brane probe computation, the enhan\c{c}on radius is now
found to be
\begin{equation}
\label{enhan2}
r_e = r_2 + 2\,\frac{N}{M}\left(r_2-r_1\right) \,.
\end{equation}
Surprisingly, when examining the relative positions of the
enhan\c{c}on and the horizon, the conclusions do not change with
respect to the pure fractional-brane case. Indeed, this follows from
a simple inspection of eqs~\eqref{r1}--\eqref{r2} which gives at hand
that $r_2-r_1$ is always positive for our solution. Thus, $r_0$ is
always smaller than $r_e$, also in the presence of  regular branes
and regardless of the relative value of $N$ with respect to $M$.
This contrasts with the result found in
refs~\cite{Johnson:1999qt,Johnson:2001wm} for branes wrapped on K3.
However, the two systems are different. In our case, as already noticed,
regular branes do not feel the enhan\c{c}on and a regular D-brane
probe can thus go all the way to the center $r=0$. Indeed, one finds
that $r_e \sim M$ at extremality not only in eq.~\eqref{enhan1} but
also in eq.~\eqref{enhan2}. The extremal $N$-regular/$M$-fractional
brane configuration can therefore be thought of as composed of $N$
regular branes at the origin and $M$ fractional branes smeared on the
enhan\c{c}on shell. In contrast, for the case discussed in
refs~\cite{Johnson:1999qt,Johnson:2001wm} the unwrapped branes (which
correspond to regular branes here) do indeed influence the enhan\c{c}on
at extremality, causing it to decrease in size. When the number of
unwrapped branes exceeds the number of wrapped ones, the enhan\c{c}on
is small enough that, for a sufficiently large non-extremality
parameter $r_0$, the system can be turned into a black hole with
$r_0$ larger than $r_e$. In our case the situation is different.
Nevertheless, one would expect a limit $N\gg M$ in which the
enhan\c{c}on at extremality would be quantitatively irrelevant and
thus should not sensibly affect the regular-brane thermodynamics. In
this respect, it would be very interesting to find the precise relation
between $r_0'$ and the external parameters when imposing thermodynamical
equilibrium between the internal black hole with horizon radius $r_0'$
and the enhan\c{c}on shell.

Having extensively discussed the possibility of our system developing a
horizon, we should also mention that there is an additional characteristic
radius whose relative position with respect to the enhan\c{c}on is of
interest. We are referring here to the radius where the potential for a
massive particle probe feeling \emph{only} the string-frame geometry becomes
repulsive. This radius (denoted $r_d$ in ref.~\cite{Johnson:2001wm}) is
readily obtained as the stationary point of the effective potential for
a radially moving probe with world-line action given by the string-frame
Nambu--Goto action. The resulting condition is
$(g_{tt}^{\mathrm{s.f.}})'(r_d)=0$, which for our non-extremal case
translates to $(f H^{-1/2})'(r_d)=0$. The solution reads (still for $p=2$)
\begin{equation}
r_d = \frac{(2q_1^2+q_2^2)(r_0-r_1)r_1 + q_2^2(r_2-r_0)r_2}{2q_1^2
(r_0-r_1)+q_2^2(r_2-r_1)} \,.
\end{equation}
It is straightforward to verify that this radius is always smaller than
or (at extremality) equal to $r_e$, and that the enhan\c{c}on hence excises
the unphysical region.

Finally, let us recall that D$p$-brane probes (fractional or regular)
moving in a background generated by the same objects also feel non-trivial
potentials out of extremality, where supersymmetry is broken. Although
these potentials are not due solely to gravity, we have nevertheless
observed that the regions where they would have become repulsive are excised
by the enhan\c{c}on. In fact, it is amusing to notice that for a
\emph{regular} D$p$-brane probing a non-extremal \emph{fractional}-brane
background, the radius at which the potential changes sign actually
coincides with the enhan\c{c}on radius.

%%%%%%%%%%%%%%%%%%%%%%%%%%%%%%%%%%%%%%%%%%%%%%%%%%%%%%%%%%%%%%%%%%%%%%%%%%%%%%%
%%%%%%%%%%%%%%%%%%%%%%%%%%%%%%%%%%%%%%%%%%%%%%%%%%%%%%%%%%%%%%%%%%%%%%%%%%%%%%%
\vspace{1ex}
\section{Discussion and conclusions}
\label{disc}

The upshot of the previous section is that a system of non-extremal
fractional branes on the orbifold $T^4\!/\ZZ_2$ cannot collapse into a
black hole. We discuss in the following a possible interpretation of
this observation in terms of the pure SYM theory living on the fractional
D-branes and the theories dual to them in the sense of the AdS/CFT
correspondence~\cite{Maldacena:1997re,Gubser:1998bc,Witten:1998qj}.

It is well known~\cite{Karch:1998yv} that a transverse T-duality on a
fractional brane gives a Hanany--Witten setup~\cite{Hanany:1997ie}.
In particular, a fractional D$p$-brane on the orbifold $T^4\!/\ZZ_2$
is  T-dual to a D($p$+1)-brane stretched between two NS5-branes, the
distance  between them being proportional to the flux $b$ of the NS-NS
two-form of  our scenario. Moreover, from the NS5-brane setup one
obtains the wrapped  brane setting of
refs~\cite{Johnson:1999qt,DiVecchia:2001uc} by a transverse
T-duality. All of these brane setups describe at low energies a pure
SYM  with eight supercharges. It was argued in
refs~\cite{Johnson:1999qt,Wijnholt:2001us} that the dual of pure SYM
with eight supercharges is a non-gravitational theory. This means that
the gravitational multiplet decouples for a fractional D-brane
solution in type-II string theory on $T^4\!/\ZZ_2$ in the decoupling
limit of the pure SYM on the brane. In the T-dual Hanany--Witten
setup this is the same as saying that gravity decouples from the
NS5-branes so that the dual theory is described by the
non-gravitational theory living on the
NS5-brane~\cite{Seiberg:1997zk,Dijkgraaf:1997ku,Aharony:1998ub}. In
the fractional-brane setup, the 5+1-dimensional fields living on the
NS5-branes correspond to the fields of the twisted sector, which
indeed have 5+1-dimensional dynamics. In fact, these are the only
fields entering all the relevant gauge-theory quantities in the
correspondence while the contribution from the gravitational
multiplet always cancels as shown for instance in
refs~\cite{Bertolini:2000dk,Grana:2001xn,Bertolini:2001qa,Billo:2001vg,
Klebanov:1999rd,Klebanov:2002gr,Bertolini:2002xu}.

Now, in the decoupling limit $r/r_e$ is fixed (this because $r\sim
\alpha'$ and $r_e\sim\alpha' \gym^2 M$ and we have $\gym^2 M$ fixed in
the limit). Hence, if there had been a horizon $r_0 > r_e$, it would
have remained after taking the decoupling limit, and we would thus
have had a black hole in the dual theory. This would have been in
contradiction with the conjecture that the dual is a non-gravitational
theory. Hence, our analysis can be seen as a further piece of evidence
that the dual theory is indeed non-gravitational. In
ref.~\cite{Johnson:1999qt} a similar  consideration was made for
D-branes wrapping K3. For that case it was found that the solution
can collapse into a black hole. However, it was subsequently shown
that the energies needed for the solution to collapse to a black hole
correspond via the gauge-theory/gravity duality to energies that are
beyond reach in the pure SYM theory. We intend to return to the
decoupling-limit issue for this kind of pure SYM
theories~\cite{Bertolini:2002ab}.

%%%%%%%%%%%%%%%%%%%%%%%%%%%%%%%%%%%%%%%%%%%%%%%%%%%%%%%%%%%%%
%%%%%%%%%%%%%%%%%%%%%%%%%%%%%%%%%%%%%%%%%%%%%%%%%%%%%%%%%%%%%
\vspace{1ex}
\section*{Acknowledgements}

We are grateful to M. Bill\'o, J. de Boer, P. Di Vecchia, E. Imeroni,
E. Kiritsis, A. Lerda, E. Lozano-Tellechea, R. Russo and M. Wijnholt
for useful discussions, and to P. Bain for some interesting remarks.
T.H. would like to thank the Niels Bohr Institute and the Institute
for Theoretical Physics of Amsterdam University, N.O. and A.W. the
Niels Bohr Institute and Nordita, and M.B. the Department for Theoretical
Physics of Turin University for hospitality while part of this work was
carried out. M.B. is supported by a European Commission Marie Curie
Postdoctoral Fellowship under contract number HPMF-CT-2000-00847.
N.O. is supported by the Stichting FOM.

%%%%%%%%%%%%%%%%%%%%%%%%%%%%%%%%%%%%%%%%%%%%%%%%%%%%%%%%%%%%%%%%%%%%%%%%%%%%
%%%%%%%%%%%%%%%%%%%%%%%%%%%%%%%%%%%%%%%%%%%%%%%%%%%%%%%%%%%%%%%%%%%%%%%%%%%%
\vspace{1ex} \appendix

\section{Details on the solutions and their derivation}
\label{sec:theapp}

In this appendix we present the derivation of the non-extremal solutions
discussed in the main text. We begin by giving the equations that they solve.

\subsection{Equations of motion}
\label{subsect:EOMs}

The equations of motion for the scalar fields encoded in the
actions~\eqref{thelagrangian} for $p=0,2,3$ and~\eqref{action1} for
$p=1$ turn out to be identical in form:
\begin{eqnarray}
\label{phieom}
\frac{1}{\sqrt{-g}} \partial_\mu ( \sqrt{-g}\,\partial^\mu\!\phi ) &=&
\frac{1-p}{4} e^{(1-p)\phi}\Big( e^\eta (G_{(p+2)})^2 +
(\tilde{F}_{(p+2)})^2 \Big) \,, \\
\label{etaeom}
\frac{1}{\sqrt{-g}} \partial_\mu ( \sqrt{-g}\,\partial^\mu\eta ) &=&
e^{(1-p)\phi} e^\eta (G_{(p+2)})^2 - e^{-\eta} \,
\partial_\mu\tilde{b}\,\partial^\mu\tilde{b} \,,\\
\label{beom}
\frac{1}{\sqrt{-g}}
 \partial_\mu(\sqrt{-g}\,e^{-\eta}\,\partial^\mu\tilde{b})  &=& -
 e^{(1-p)\phi} \,G_{(p+2)}\cdot \tilde{F}_{(p+2)} \,.
\end{eqnarray}
Here we have introduced the notation  $(G_{(n)})^2 \equiv G_{(n)}\cdot
G_{(n)}\equiv \frac{1}{n!} G_{\mu_1  \ldots \mu_{n}} G^{\mu_1 \ldots
\mu_{n}}$.

For the gauge fields $A_{(p+1)}$ and $C_{(p+1)}$ the case $p=1$
differs slightly from the others; while for $p=0,2,3$, the equations
of motion  all take the common form
\begin{eqnarray}
\label{Aeom}
\partial_{\mu_1} (\sqrt{-g} \, e^{(1-p)\phi} \,  \tilde{F}^{\mu_1
 \ldots \mu_{p+2}}) &=& 0\,, \\
\label{Ceom}
\partial_{\mu_1} ( \sqrt{-g} \, e^{(1-p)\phi} \,e^\eta \,
 \tilde{G}^{\mu_1 \ldots \mu_{p+2}} ) &=& 0 \,,
\end{eqnarray}
we obtain instead for $p=1$ (in form notation)
\begin{eqnarray}
\label{Aeom1}
d\,{*}\Big(\tilde{F}_{(3)}+\tilde{b}\,{*}G_{(3)}\Big) &=& 0 \,, \\[1ex]
\label{SDcond}
\tilde{F}_{(3)} - {*}\tilde{F}_{(3)} &=& 0 \,, \\[1ex]
\label{Ceom1}
d\,{*}\Big(e^\eta\,\tilde{G}_{(3)}+\frac{1}{2}\,
   \tilde{b}^2\,{*}G_{(3)}\Big) &=& 0 \,.
\end{eqnarray}
Here we included the self-duality condition for $\tilde{F}_{(3)}$, as
this constraint on the solution has the status of an equation of
motion.%
\footnote{Eqs~\eqref{Aeom1} and~\eqref{SDcond} are not independent;
the self-duality condition is stronger and implies~\eqref{Aeom1}.}
For convenience we have introduced the notation
\begin{equation}
\tilde{G}_{(p+2)} = G_{(p+2)}-
  e^{-\eta}\,\tilde{b}\,\tilde{F}_{(p+2)}\,.
\end{equation}

To the above equations should be added the Einstein equations which we
shall give once we have introduced the spherically symmetric ansatz
that we will employ. Let us first, however, mention that under such an
ansatz the self-dual fractional D1-brane case can alternatively be
solved by taking a standard electric field-strength ansatz for
$\tilde{F}_{(3)}$ and using the equations of motion obtained for
$p=1$ from the ``naive'' action~\eqref{thelagrangian}. The dual,
magnetic, components of $\tilde{F}_{(3)}$ (as well as the full
potential $A_{(2)}$), if required, can be obtained by imposing the
self-duality condition at the very end. We will adopt this (standard)
effective procedure below, allowing us to discuss the case $p=1$ in
parallel with the others.

%%%%%%%%%%%%%%%%%%%%%%%%%%%%%%%%%%%%%%%%%%%%%%%%%%%%%%%%%%%%%%%%%%%%%
\subsection{The spherical $p$-brane ansatz}
\label{subsect:spherical}
A general ansatz for a metric possessing the symmetries of the
non-extremal solution is
\begin{equation}
\label{genansatz}
ds^2 = - B^2 dt^2 + C^2 \sum_{i=1}^p (dx^i)^2 + F^2 dr^2 + G^2 r^2
d\Omega_{4-p}^2 \,,
\end{equation}
where $B$, $C$, $F$ and $G$ are functions of the transverse radial
coordinate $r$ only. The equations of motion~\eqref{Aeom}--\eqref{Ceom}
for the two gauge fields $A_{(p+1)}$ and $C_{(p+1)}$ can trivially be
integrated to give%
\footnote{In form notation this corresponds to $\tilde{F}_{(p+2)}
=(-1)^{p}\qh_1e^{-(1-p)\phi}\,{*}d\Omega_{4-p}$, so that
${*}d{*}[e^{(1-p)\phi}\tilde{F}_{(p+2)}]= -\qh_1{*}d^2\Omega\equiv0$
and  $\int_{S^{4-p}}e^{(1-p)\phi}\,{*}\tilde{F}_{(p+2)}=  -
\qh_1\,\Omega_{4-p}$.}
\begin{eqnarray}
\tilde{F}_{r 0 \cdots p} &=& e^{-(1-p)\phi} B C^p F
       \frac{\qh_1}{(Gr)^{4-p}}\,, \\[2ex] \tilde{G}_{r 0 \cdots p}
       &=& e^{-(1-p)\phi} e^{-\eta} B C^p F \frac{\qh_2}{(Gr)^{4-p}}
       \,,
\end{eqnarray}
where we used the result $\sqrt{-g} = \sqrt{g_{\Omega_{4-p}}} r^{4-p}
B C^p F G^{4-p}$. The two ``hatted'' charges, $\qh_1$ and $\qh_2$,
introduced as a  shorthand notation for this appendix, each absorbs a
$p$-dependent factor according to
\begin{equation}
\label{appcharges}
\qh_1 = k_p\,q_1 \spa \qh_2 = k_p\,q_2 \spa  k_p =
\left\{\begin{array}{cll} 3-p\,, &~~&p=0,1,2 \\[.5ex]  1\,, & & p=3
\end{array}\right. \,,
\end{equation}
with $q_1$ and $q_2$ being the charges which appear in the solutions
and which are natural from the physical perspective. Defining
\begin{equation}
\hat{b} = 1 + \frac{q_1}{q_2} \tilde{b} \,,
\end{equation}
we have
\begin{equation}
\label{Gspherical}
G_{r 0 \cdots p} = e^{-(1-p)\phi}\, e^{-\eta} B C^p F
\frac{\qh_2\,\hat{b}}{(Gr)^{4-p}} \,,
\end{equation}
so that
\begin{eqnarray}
e^{(1-p)\phi} \,e^\eta (G_{(p+2)})^2 &=& - e^{-(1-p)\phi}\,
e^{-\eta}\, \hat{b}^2 \frac{\qh_2^2}{(Gr)^{2(4-p)}} \,, \\[2ex]
e^{(1-p)\phi} (\tilde{F}_{(p+2)})^2 &=& - e^{-(1-p)\phi}
\frac{\qh_1^2}{(Gr)^{2(4-p)}} \,.
\end{eqnarray}
Introducing the notation
\begin{equation}
L \equiv B C^p F^{-1} (Gr)^{4-p} \,,
\end{equation}
\begin{equation}
Y \equiv -e^{(1-p)\phi} \Big( e^\eta (G_{(p+2)})^2 +
 (\tilde{F}_{(p+2)})^2 \Big) = \frac{e^{-(1-p)\phi}}{(Gr)^{2(4-p)}}
 \Big( \qh_1^2 + \qh_2^2 e^{-\eta} \hat{b}^2 \Big) \,,
\end{equation}
the scalar-field equations of motion~\eqref{phieom}--\eqref{beom} can
be written compactly as
\begin{eqnarray}
\label{dilequ}
\phi'' + \phi' (\log L)' &=& - \frac{1-p}{4} F^2 Y \,, \\
\label{etaequ}
\eta'' + \eta' (\log L)' &=& - F^2 e^{-(1-p)\phi} e^{-\eta} \hat{b}^2
\frac{\qh_2^2}{(Gr)^{2(4-p)}} - \frac{q_2^2}{q_1^2}e^{-\eta}
(\hat{b}')^2 \,, \\
\label{bequ}
L^{-1} (L\,e^{-\eta}\,\hat{b}')' &=& F^2 e^{-(1-p)\phi} e^{-\eta}
\frac{\qh_1^2}{(Gr)^{2(4-p)}} \hat{b} \,.
\end{eqnarray}
Finally, the Einstein equations read (no summation over indices)
\begin{equation}
\label{Rttequ}
R^t_{\ t} = \frac{p-3}{8} Y \,,
\end{equation}
\begin{equation}
\label{Riiequ}
R^i_{\ i} = \frac{p-3}{8} Y \,,
\end{equation}
\begin{equation}
\label{Rrrequ}
R^r_{\ r} = F^{-2} \left( (\phi')^2 + \frac{1}{4} (\eta')^2 +
\frac{1}{2} \frac{q_2^2}{q_1^2} e^{-\eta} (\hat{b}' )^2 \right) +
\frac{p-3}{8} Y \,,
\end{equation}
\begin{equation}
\label{Raaequ}
R^\alpha_{\ \alpha} = \frac{p+1}{8} Y \,.
\end{equation}
Here (for $p>0$) $i=1,\ldots,p$ are the spatial world-volume directions
while $\alpha$ runs over the $4-p$ transverse angular coordinates. The
Ricci-tensor components for a metric of the form~\eqref{genansatz} are
\begin{equation}
R^{t}_{\ t} = \frac{1}{F^2} \Big(-(\log B)''-(\log B)'(\log L)'\Big)
\,,\\[2ex]
\end{equation}
\begin{equation}
R^{i}_{\ i} = \frac{1}{F^2} \Big( - (\log C)'' - (\log C)' (\log L)'
\Big) \,,
\end{equation}
\begin{eqnarray}
R^{r}_{\ r} &=& \frac{1}{F^2} \Big( -(\log B)'' - ((\log B)')^2
   +(\log F)'\,(\log B)' \nn \\ & & \quad +\,p\,\Big[-(\log C)''
   -((\log C)')^2 + (\log F)'\,(\log C)' \Big] \nn \\ & & \quad
   +\,(4-p)\,\Big[ -(\log Gr)'' -((\log Gr)')^2
   + (\log F)'\,(\log Gr)'\Big] \Big) \,,
\end{eqnarray}
\begin{equation}
\label{Raa}
R^{\alpha}_{\ \alpha} = \frac{1}{F^2} \Big( -(\log Gr)''
- (\log Gr)'\,(\log L)' + (3-p) \frac{F^2}{(Gr)^2} \Big) \,.
\end{equation}
%

%%%%%%%%%%%%%%%%%%%%%%%%%%%%%%%%%%%%%%%%%%%%%%%%%%%%%%%%%%%%%%%%%%%
\subsection{Solving the equations of motion}
\label{subsect:solving}

In order to find the non-extremal versions of the supergravity
solutions for fractional D0- and D2-branes on $T^4\!/\ZZ_2$ of
ref.~\cite{Frau:2000gk} that reduce to Schwarz\-schild black-brane
metrics for vanishing charges, the natural ansatz to employ is
\begin{equation}
\label{our_ansatz}
ds^2 = H^{\frac{p-3}{4}} \Big( - f dt^2 + \sum_{i=1}^p (dx^i)^2 \Big)
    + H^{\frac{p+1}{4}} \left( f^{-1}dr^2 + r^2 d\Omega_{4-p}^2
    \right)\,.
\end{equation}
This ansatz applies equally well for $p=1$, while for the three-brane,
as mentioned in section~\ref{sect:solutions}, it is preferable to use
the adapted ansatz
\begin{equation}
\label{our_ansatz3}
ds^2 = - f dt^2 + \sum_{i=1}^3 (dx^i)^2 +
 \Big(\frac{r_\Lambda}{r}\Big)^2 H \left( f^{-1}dr^2 + r^2 d\theta^2
 \right) \,.
\end{equation}
From these metrics we can read off expressions for $B$, $C$, $F$ and
$G$ in terms of the radial functions $H$ and $f$, and plug them in the
equations of motion listed above. Solving the so obtained equations is
the objective of the present section.

We start by observing that the harmonic equation~\eqref{eq:f_eq} for
$f$ follows from the Einstein equations~\eqref{Rttequ}
and~\eqref{Riiequ} (for $p=0$, eq.~\eqref{Riiequ} is not present and
other equations need to be used instead). Equipped with this result,
we combine the dilaton equation~\eqref{dilequ} with the angular
Einstein equation~\eqref{Raaequ} to get the equation
\begin{equation}
\label{chieq}
\chi'' + \chi' (\log L)' = 0 \,,
\end{equation}
where we have defined $\chi = \log(e^\phi H^{-(1-p)/4})$ and where now
$L=fr^{4-p}$. Requiring that the dilaton behave in a non-singular
manner at the horizon we find that $\chi$ must be constant. Asking
furthermore that at infinity we recover flat Minkowski space with
vanishing dilaton, we obtain
\begin{equation}
\label{phiHrelp}
e^\phi = H^{\frac{1-p}{4}} \,.
\end{equation}
For the case $p=1$, eq.~\eqref{chieq} simply gives that the dilaton is
constant, as appropriate for the string in six dimensions.

It remains to solve the scalar-field equations
\eqref{etaequ}--\eqref{bequ} together with the radial Einstein
equation~\eqref{Rrrequ}, using also either~\eqref{dilequ}
or~\eqref{Rttequ}. With
\begin{equation}
\label{F2Yp}
F^2 Y = \frac{1}{H f r^{2(4-p)}} \Big( \qh_1^2+\qh_2^2 e^{-\eta}
\hat{b}^2 \Big)\,,
\end{equation}
the two latter amount to the single equation
\begin{equation}
\label{newdilequ} e^{-\eta} \hat{b}^2 = \gamma \,,
\end{equation}
where we have introduced
\begin{equation}
\gamma \equiv -\frac{q_1^2}{q_2^2} - \frac{1}{\qh_2^2} H r^{4-p}
\left( f r^{4-p} \frac{H'}{H} \right)' \,.
\end{equation}
From \eqref{etaequ} and~\eqref{bequ} we, on the other hand, obtain the
equation
\begin{equation}
\left( L \eta' + \frac{q_2^2}{q_1^2} L e^{-\eta} \hat{b} \, \hat{b}'
\right)' = 0 \,,
\end{equation}
which can trivially be integrated to
\begin{equation}
\label{q3eq} \left( e^\eta + \frac{1}{2} \frac{q_2^2}{q_1^2}
\hat{b}^2 \right)' = q_3 L^{-1} e^\eta \,,
\end{equation}
$q_3$ being a constant of integration. Since $L = f r^{4-p}$ we see
that the right-hand side becomes singular at the horizon unless
$q_3=0$. Requiring the scalars $\eta$ and $\hat{b}$ to be non-singular
at the horizon we are hence led to set $q_3$ to zero. Imposing,
furthermore, that these scalars vanish at infinity we arrive at the
relation
\begin{equation}
e^\eta = 1 + \frac{1}{2} \frac{q_2^2}{q_1^2} - \frac{1}{2}
\frac{q_2^2}{q_1^2}\, \hat{b}^2 \,,
\end{equation}
which when combined with the dilaton equation written in the
form~\eqref{newdilequ} finally yields the expressions
\begin{eqnarray}
e^{\eta} &=& \left( 1 + \frac{1}{2} \frac{q_2^2}{q_1^2} \right) \left[
1 + \frac{1}{2} \frac{q_2^2}{q_1^2} \gamma \right]^{-1} \,, \\[2ex]
\hat{b} &=& \sqrt{\gamma} \left( 1 + \frac{1}{2} \frac{q_2^2}{q_1^2}
\right)^{1/2} \left[ 1 + \frac{1}{2} \frac{q_2^2}{q_1^2} \gamma
\right]^{-1/2} \,.
\end{eqnarray}
At this point, it remains only to determine the function $H$ by
solving the equations~\eqref{etaequ} and~\eqref{Rrrequ} with the above
results as input. To this end, we define the functions $h_1$ and $h_2$
by
\begin{equation}
\label{h1h2defs}
\gamma = \frac{h_2^2}{H} \spa \frac{1+\frac{1}{2} \frac{q_2^2}{q_1^2}
\gamma }{1 +\frac{1}{2} \frac{q_2^2}{q_1^2}} = \frac{h_1^2}{H} \,,
\end{equation}
so that
\begin{equation}
\label{HetaDh1h2}
H = \left( 1 + \frac{1}{2} \frac{q_2^2}{q_1^2} \right) h_1^2 -
\frac{1}{2} \frac{q_2^2}{q_1^2}\, h_2^2 \spa e^\eta = \frac{H}{h_1^2}
\spa \hat{b} = \frac{h_2}{h_1} \,.
\end{equation}
Using these expressions the gauge field strengths take the form
\begin{equation}
\label{fstrengthsolsapp}
G_{r0\ldots p} = \frac{k_p\,q_2}{r^{4-p}} \frac{h_1 h_2}{H^2} \spa
\tilde{G}_{r0\ldots p} = \frac{k_p\,q_2}{r^{4-p}} \frac{h_1^2}{H^2}
\spa \tilde{F}_{r0\ldots p} = \frac{k_p\,q_1}{r^{4-p}} \frac{1}{H} \,.
\end{equation}
Having expressed all fields in terms of $h_1$ and $h_2$, we rewrite
also the remaining equations of motion which determine these
functions. Equation~\eqref{Rrrequ} thus reads
\begin{equation}
\label{Rrrh1h2}
\left( 1 + \frac{1}{2}\frac{q_2^2}{q_1^2} \right) h_1 \left( h_1'' +
\frac{4-p}{r} h_1' \right) = \frac{1}{2}\frac{q_2^2}{q_1^2} h_2 \left(
h_2'' + \frac{4-p}{r} h_2' \right) \,,
\end{equation}
while \eqref{etaequ} may be written as
\begin{eqnarray}
\label{etah1h2}
0 &=& \frac{\qh_1^2}{r^{2(4-p)}} + H f
 \Big(\log\frac{H}{f}\Big)'\,\Big(\log\frac{h_2}{h_1}\Big)'
 \nonumber\\[1ex] && +\, H f
 \left[\frac{1}{h_1}\left(h_1''+\frac{4-p}{r}h_1'\right) -
 \frac{1}{h_2}\left(h_2'' + \frac{4-p}{r} h_2' \right) \right] \,.
\end{eqnarray}
Clearly, the simplest non-trivial way to satisfy eq.~\eqref{Rrrh1h2}
is by taking both $h_1$ and $h_2$ to be harmonic, like $f$, and we
will do so here.  Taking the boundary conditions at infinity into
account we thus have for  $p=0,1,2$
\begin{equation}
\label{theh1h2}
f = 1 - \frac{r_0^{3-p}}{r^{3-p}} \spa h_1 = 1 -
  \frac{r_1^{3-p}}{r^{3-p}} \spa h_2 = 1-\frac{r_2^{3-p}}{r^{3-p}} \,.
\end{equation}
For $p=3$ the three harmonic functions governing the non-extremal
solution instead take the form
\begin{equation}
\label{theh1h23}
f= 1 - a_0 \log\frac{r_\Lambda}{r} \spa h_1 = 1 - a_1 \log
\frac{r_\Lambda}{r} \spa h_2 = 1 - a_2 \log \frac{r_\Lambda}{r}\,,
\end{equation}
where $r_\Lambda$ is a large-radius cut-off.  In both cases, we are
left with two undetermined parameters---$r_{1,2}$ and $a_{1,2}$
respectively---and in both cases these parameters are fixed by
eq.~\eqref{etah1h2}. However, before analysing this equation, let us
derive the solutions for the gauge potentials $C_{(p+1)}$ and
$A_{(p+1)}$.

Starting with the untwisted sector, we get an integral for $C_{0\ldots
p}$ from eq.~\eqref{fstrengthsolsapp} by recalling that $G_{r0\ldots
p} = C_{0\ldots p}'$. For $p=0,1,2$, this integral can readily be
evaluated with the result
\begin{equation}
\label{Cpotsol}
C_{0\ldots p} = - \frac{q_2}{r^{3-p}} \frac{h_3}{H} \spa h_3 \equiv
 \frac{1}{2}(h_1+h_2) \ .
\end{equation}
To obtain this result we used the fact that $h_1$ and $h_2$ being
harmonic implies the identity
\begin{equation}
\label{h3DE}
h_3 H + \frac{r}{3-p} ( h_3 H' - h_3' H ) = h_1 h_2 \,.
\end{equation}
With $C_{0\ldots p}$ in hand, we similarly obtain $A_{0\ldots p}$ by
integrating the equation
\begin{equation}
\label{Adiffeq}
A_{0\ldots p}' = \tilde{F}_{r0\ldots p} - \tilde{b}'\,C_{0\ldots p} =
 (3-p)\frac{q_1}{r^{4-p}} \frac{1}{H} \left[ 1 + \frac{r}{3-p}
 \frac{q_2^2}{q_1^2} h_3 \left( \frac{h_2}{h_1} \right)' \right] \,,
\end{equation}
The solution is found to be
\begin{equation}
\label{Apotsol}
A_{0\ldots p} = - \frac{q_1}{r^{3-p}}\frac{1}{h_1} \,.
\end{equation}
Here we used the identity
\begin{equation}
\label{h3DE2}
H = h_1^2 + \frac{r}{3-p}\frac{q_2^2}{q_1^2}h_3(h_1 h_2' - h_1' h_2)
\,,
\end{equation}
which, like~\eqref{h3DE}, follows directly from the form of $H$ and
$h_3$ in terms of $h_1$ and $h_2$.

For the three-brane the situation is somewhat more subtle since
there are no analogues of the first-order differential
equations~\eqref{h3DE} and~\eqref{h3DE2}. Instead, led by the
form of the harmonic functions, we apply the mapping
$r^{-(3-p)}\mapsto\log(r_\Lambda/r)$ to the potentials
in~\eqref{Cpotsol} and~\eqref{Apotsol} above. It is then
straightforward to check that the so obtained expressions
give the correct gauge potentials:
\begin{equation}
C_{0123} = - \frac{q_2\,h_3}{H}\,\log\frac{r_\Lambda}{r} \spa
A_{0123} = - \frac{q_1}{h_1}\,\log\frac{r_\Lambda}{r} \,.
\end{equation}

Let us then, finally, address eq.~\eqref{etah1h2}, which for
harmonic $h_1$ and $h_2$ immediately simplifies to
\begin{equation}
\label{etah1h2harm}
 \frac{(k_p\,q_1)^2}{r^{2(4-p)}} + H f
  \Big(\!\log\frac{H}{f}\Big)'\,\Big(\!\log\frac{h_2}{h_1}\Big)' =
  0\,.\\[1ex]
\end{equation}
By performing the $p$-dependent substitutions of variables and
parameters
\begin{equation}
\sigma = \left\{\begin{array}{lll} r^{-(3-p)}\, &~~& p=0,1,2 \\[1ex]
     \log{\frac{r_\Lambda}{r}}  & & p=3 \end{array} \right. \spa\quad
\rho_{0,1,2} = \left\{\begin{array}{lll} r_{0,1,2}^{3-p} &~~& p=0,1,2
  \\[1ex] a_{0,1,2}   & & p=3 \end{array} \right. \,,
\end{equation} 
this equation reduces to the single, \emph{$p$-independent}, equation
\begin{equation}
\label{singleeq}
q_1^2 + \hH \hf\Bigg(\!\log\frac{\hH}{\hf}\Bigg)'
         \Bigg(\!\log{\frac{\hh_2}{\hh_1}}\Bigg)'=0 \,,
\end{equation}
where
\begin{equation}
\label{genharm}
\hf=1-\rho_0\,\sigma \spa  \hh_{1,2}=1-\rho_{1,2}\,\sigma \spa \hH =
\left( 1 + \frac{1}{2} \frac{q_2^2}{q_1^2} \right) \hh_1^2 -
\frac{1}{2} \frac{q_2^2}{q_1^2}\,\hh_2^2 \,,
\end{equation}
and the prime now denotes differentiation with respect to $\sigma$.
Note that only the ``un-hatted'' charges $q_1$ and $q_2$ enter in
eq.~\eqref{singleeq}, showing that these are the charges that appear
in the harmonic functions. Inserting the expressions~\eqref{genharm}
we find that the two undetermined parameters $\rho_1$ and $\rho_2$
are required by~\eqref{singleeq} to satisfy
\begin{equation}
\label{r1r2rel}
\rho_1 = u \spa \rho_2 = \frac{ u \rho_0 }{2u - \rho_0} \,,
\end{equation}
where $u$ is a solution of the quartic equation
\begin{eqnarray}
\label{qua1}
 4\,(2\,q_1^2+q_2^2)\,u^4 - 8\,(2\,q_1^2 + q_2^2)\,\rho_0\,u^3 &+&
  2\,(2\,q_2^2\, \rho_0^2 + 5\, q_1^2\,\rho_0^2 - 2\,q_1^4 )\,u^2
  \mbox{~~~~~~~~~~~~~~~} \nn \\ &+& 2\,(2 \,q_1^4 \,\rho_0 - q_1^2\,
  \rho_0^3)\, u - q_1^4 \,\rho_0^2 = 0 \,.
\end{eqnarray}
This equation has the four solutions
\begin{equation}
u =\frac{1}{2} \rho_0 + \epsilon_1 \frac{\sqrt{2q_1^4 + (q_1^2 +
q_2^2)\rho_0^2 -\epsilon_2 2 q_1^2 \Lambda}}{2 \sqrt{2q_1^2+q_2^2}} \,,
\end{equation}
where $\epsilon_1 = \pm 1$, $\epsilon_2 = \pm 1$ and
\begin{equation}
\Lambda = \sqrt{q_1^4 + (q_1^2 +q_2^2)\rho_0^2 + \frac{1}{4}\rho_0^4}
\;.
\end{equation}
Consequently, the solutions to the equations of motion have four
branches given by
\begin{eqnarray}
\label{r1_app}
\rho_1 &=& \frac{1}{2} \rho_0 + \epsilon_1 \frac{\sqrt{2q_1^4 + (q_1^2
+ q_2^2)\rho_0^2 -\epsilon_2 2 q_1^2
\Lambda}}{2\sqrt{2q_1^2+q_2^2}}\,, \\[2ex]
\label{r2_app}
\rho_2 &=& \frac{1}{2} \rho_0 + \epsilon_1 \frac{\sqrt{2q_1^4 + (q_1^2
+ q_2^2)\rho_0^2 + \epsilon_2 2 q_1^2 \Lambda}}{2\sqrt{q_2^2}} \,,
\end{eqnarray}
where, again, $\epsilon_1 = \pm 1$ and $\epsilon_2 = \pm 1$.  There
are thus four solutions, which may be summarized by the expressions
given  in section \ref{subsect:ansaetze}, with $\rho_{1,2}$ as in
\eqref{r1_app}, \eqref{r2_app}. By taking the limit $\rho_0
\rightarrow 0$ in the these expressions, one can easily see that the
branch which correctly reduces to the extremal fractional-brane
solution---i.e.\ the branch which satisfies the boundary conditions
imposed by the action~\eqref{bac1}---and which thus represents the
non-extremal fractional brane solution, is the branch with $\epsilon_1
=+1$ and $\epsilon_2 =+1$. This choice corresponds
to~\eqref{r1}--\eqref{r2} ($p=0,1,2$) and~\eqref{a1}--\eqref{a2} 
($p=3$) in the text.

\subsection{Some properties of the four branches}
\label{subsect:iquattrorami}

To understand the properties of these four branches it is useful to
consider the sign of $H$ at the horizon. (Since, as pointed out in
section~\ref{subsect:ansaetze}, the three-brane case is special, we
restrict the discussion to $p<3$.) Thus, setting $r=r_0$ in
eq.~\eqref{etah1h2harm} we find that
\begin{equation}
H(r_0) = (3-p)\,\frac{q_1^2}{r_0^{7-2p}} \left( \frac{h_2'}{h_2} -
\frac{h_1'}{h_1} \right)^{-1} \Bigg\vert_{r=r_0} \,.
\end{equation}
Using eq.~\eqref{theh1h2} we then obtain
\begin{equation}
\label{signH} \mathrm{sign} (H(r_0)) = \mathrm{sign}(r_2-r_1) \;
\mathrm{sign}(r_0-r_1) \; \mathrm{sign}(r_0-r_2) \,.
\end{equation}
Hence, $H(r_0)$ is positive if either $r_0 < r_1 < r_2$ or $r_1 < r_2
< r_0$ or $r_2 < r_0 < r_1$, while it is negative if either $r_1 < r_0
< r_2$ or $r_0 < r_2 < r_1$ or $r_2 < r_1 < r_0$.

One can now check that the sign of $H(r_0)$ cannot be changed within a
particular branch. From eq.~\eqref{signH} we see that this precisely
means that $r_0$, $r_1$ and $r_2$ cannot cross within a particular
branch. In table~\ref{tabbranches} we have listed the restrictions on
the ranges for $r_0$, $r_1$ and $r_2$, along with the sign of $H(r_0)$
for the four branches. The solution discussed in the text corresponds
to branch I.

\begin{table}
\begin{center}
\begin{tabular}{|c|r|r|c|c|}
\hline branch & $\epsilon_1$ & $\epsilon_2$ & restrictions &
$\mathrm{sign}(H(r_0))$\\ \hline {\rm I} & $+1$ & $+1$ & $r_1 < r_0 <
r_2 $ & $-$ \\ {\rm II} & $-1$ & $+1$ & $r_2 < r_1 < r_0 $ & $-$ \\
{\rm III} & $+1$ & $-1$ & $r_2 < r_0 < r_1 $ & $+$ \\ {\rm IV} & $-1$
& $-1$ & $r_1 < r_2 < r_0 $ & $+$ \\ \hline
\end{tabular}
\caption{Properties of the four branches.\label{tabbranches}}
\end{center}
\end{table}

In order to further examine the physics of the four branches obtained
above we compute the ADM mass (per unit $p$-volume)
\begin{equation}
M_p = \frac{\Omega_{4-p}}{16\pi G_6 }\Big[ (4-p) r_0^{3-p} + (3-p) \xi
\Big]  \spa 16 \pi G_6 = \frac{2 \kappa^2}{V}
\end{equation}
where $\Omega_{4-p}$ denotes the volume of the unit $(4-p)$-sphere and
\begin{equation}
\label{xidef}
\xi = \frac{1}{q_1^2}\left[ -( 2 q_1^2 + q_2^2) r_1^{3-p} + q_2^2
  r_2^{3-p}  \right]
\end{equation}
is the coefficient of the leading $1/r^{3-p}$ term in the function
$H$ in eq.~\eqref{HetaDh1h2}. Focusing first on the four branches at
$r_0=0$, it is not difficult to see that branches~II and~III are
unphysical since they both have negative mass, and hence should be
discarded. On the other hand, for branches~I and~IV we find
\begin{equation}
\label{spxi}
\xi = \left\{ \begin{array}{ccc} q_2\,, &~~& {\rm(I)} \\[1ex]
\sqrt{2q_1^2 +q_2^2}\,, & &{\rm(IV)} \end{array} \right.
\end{equation}
so that branch~I at $r_0=0$ has the lowest mass.
To obtain \eqref{spxi} we have used~\eqref{exth1h2} for branch I while 
for branch IV at $r_0=0$  we have
$ h_1 = 1 + q_1^2/(\sqrt{2 q_1^2 + q_2^2} r^{3-p} )$, $h_2 =1$.
The charges are
\begin{eqnarray}
\label{Q1}
\hat{Q}_1 &=& \frac{- 1}{16\pi G_6 } \int_{S^{4-p}} e^{(1-p)\phi}\,
    {*}\tilde{F}_{(p+2)} = (3-p)\, q_1\,\frac{\Omega_{4-p}}{16\pi G_6 } \,,
        \\[1ex]
\label{Q2}
\hat{Q}_2 &=& \frac{- 1}{16\pi G_6 } \int_{S^{4-p}} e^{(1-p)\phi}\,e^\eta\,
  {*}\tilde{G}_{(p+2)} = (3-p)\,q_2\,\frac{\Omega_{4-p}}{16\pi G_6 } \,,
\end{eqnarray}
(recall that $16\pi G_6 = 2\kappa^2/V$).
Since the untwisted charge is $\hat{Q}_2$ irrespective of the branch,
we conclude that branch~I at extremality is BPS ($M_p = \hat Q_2$), while
branch~IV apparently describes a system that is non-BPS even in the limit
$r_0 = 0$.

Although it is presently not clear what the precise physical meaning, if
any, of the supergravity solution of branch~IV is, we note that this solution
has well-defined black-brane thermodynamics. Using standard methods of
black-hole thermodynamics, we compute the temperature and entropy:%
\footnote{We note that the following expressions are algebraically valid
for all branches. However, we have already excluded branches~II and~III
on the grounds of positivity of the mass, while for branch~I it is seen from
table~\ref{tabbranches} that $H(r_0) < 0$ and hence the expressions below
are not physically meaningful. Alternatively, we have already argued
extensively in the text that branch~I does not develop into a black brane.}
\begin{equation}
T = \frac{3-p}{4\pi} \frac{1}{r_0\sqrt{H(r_0)}} \spa
S = \frac{\Omega_{4-p}}{4 G_6 } r_0^{4-p} \sqrt{H(r_0)} \ .
\end{equation}
Using the Wess--Zumino term of the world-volume action~\eqref{bac1}
the corresponding chemical potentials, dual to the charges in~\eqref{Q1}
and~\eqref{Q2}, are
\begin{equation}
\mu_1 = - (A_{01\ldots p} + \tilde{b}\,C_{0 1 \ldots p}) \Big\vert_{r=r_0} \spa
\mu_2 = - C_{0 1 \ldots p} \Big\vert_{r=r_0} \,.
\end{equation}
More explicitly, using~\eqref{HetaDh1h2}, \eqref{Cpotsol}
and~\eqref{Apotsol} the chemical potentials read    
\begin{equation}
\mu_1 = \frac{q_1}{r_0^{3-p}} \frac{1}{h_1(r_0)} + \frac{q_2}{q_1} \left(
\frac{h_2(r_0)}{h_1(r_0)} -1 \right)
\frac{q_2}{r_0^{3-p}} \frac{h_3(r_0)}{H(r_0)} \spa
\mu_2 = \frac{q_2}{r_0^{3-p}} \frac{h_3(r_0)}{H(r_0)} \,,
\end{equation}
in terms of the harmonic functions $h_i$, $i=1,2,3$ given in
eqs~\eqref{theh1h2} and~\eqref{Cpotsol}.

As a check, we note that the first law of thermodynamics
$d M_p = T d S + \mu_i d \hat{Q}_i$ can be integrated to yield Smarr's formula
\begin{equation}
(3-p) M_p = (4-p) TS + (3-p)(\mu_1 \hat{Q}_1 + \mu_2 \hat{Q}_2) \,.
\end{equation}
One may verify explicitly that this law holds since
\begin{equation}
\label{xiid}
\xi = \mu_1 q_1 + \mu_2 q_2 \,,
\end{equation}
where $\xi $ is defined in \eqref{xidef}. To prove~\eqref{xiid} one
uses the non-trivial identity
\begin{equation}
\label{ntid}
\frac{1}{r_0^{3-p}} \frac{1}{h_1(r_0)} \left[ q_1^2 + q_2^2
\frac{h_2(r_0) h_3 (r_0)}{H(r_0)} \right] = \xi \,.
\end{equation}
which holds for all branches. To verify this statement one uses the
form of $H$ in~\eqref{HetaDh1h2} along with $h_3$ in terms of $h_{1,2}$,
and subsequently employs the relation~\eqref{r1r2rel} to eliminate $r_2$.
The identity then reduces exactly to the quartic equation~\eqref{qua1}
satisfied by $r_1$.

%%%%%%%%%%%%%%%%%%%%%%%%%%%%%%%%%%%%%%%%%%%%%%%%%%%%%%%%%%%%%%%%%%%%%%%%

\vspace{1ex}

\addcontentsline{toc}{section}{\numberline{}References}

\providecommand{\href}[2]{#2}\begingroup\raggedright\endgroup

\end{document}